# Conduction of topologically-protected charged ferroelectric domain walls

Weida Wu[1*], Y. Horibe[1], N. Lee[1], S.-W. Cheong[1] and J.R. Guest[2]

[1]*Department of Physics and Astronomy and Rutgers Center for Emergent Materials, Rutgers University, Piscataway, NJ 08854 USA*
[2]*Center for Nanoscale Materials, Argonne National Laboratory, Argonne, Illinois 60439 USA*

**Abstract**

We report on the observation of nanoscale conduction at ferroelectric domain walls in hexagonal HoMnO$_3$ protected by the topology of multiferroic vortices using *in situ* conductive atomic force microscopy, piezoresponse force microscopy, and kelvin-probe force microscopy at low temperatures. In addition to previously observed Schottky-like rectification at low bias [Phys. Rev. Lett., **104**, 217601 (2010)], conductance spectra reveal that negatively charged tail-to-tail walls exhibit enhanced conduction at high forward bias, while positively charged head-to-head walls exhibit suppressed conduction at high reverse bias. Our results pave the way for understanding the semiconducting properties of the domains and domain walls in small-gap ferroelectrics.

Topology is the foundation for numerous robust phenomena such as topological defects [1], quantum hall effect [2] and topological insulators [3] in condensed matter physics because it is insensitive to continuous deformation or perturbation. Topological defects are pervasive in complex matter such as superfluid, liquid crystals and the early universe [1, 4]. However, little is known about topological defects in systems with coupled order parameters such as multiferroics, where magnetism and ferroelectricity coexist [5, 6]. A unique kind of topological defect, where six interlocked structural antiphase and ferroelectric domain walls merge into a vortex core, was unambiguously identified in multiferroic hexagonal YMnO$_3$ [7]. The intriguing vortex-antivortex network pattern has elegantly been analyzed using graph coloring theory [8]. The band gap of hexagonal (*h*-) *RE*MnO$_3$ (*RE* = Sc, Y, Ho, ... Lu) is relatively small (~1.7 eV) [9], so the conduction properties of ferroelectric vortices may show interesting behavior. In fact, a significant conduction difference between opposite-polarization "domains" has been observed, and this conduction contrast was shown to originate from a polarization modulated Schottky-like barrier [7, 10]. Interestingly, the topology of vortex-antivortex network in 3-dimensional crystal naturally results in the presence of numerous "charged ferroelectric domain walls" where opposite polarizations are facing to each other [8, 11, 12]. Thus, it is imperative to find out the conduction properties of these charged domain walls.

Domain walls are kink solitons that separate domains with different orientations of ferroic order. They may host emergent phases or novel properties such as local conduction in an insulating matrix [13-15]. In conventional ferroelectrics, most studied domain walls are neutral while charged domain walls are rarely observed, which is likely due to unfavorable electrostatic and/or strain energy cost [16, 17]. Experimentally, observed charged domain walls are often associated with defects or needle-shape domains during polarization reversal [18-20]. However, it has been predicted that charged domain walls may be stabilized by charged defects or free charge carriers [21-23]. In contrast to conventional ferroelectrics, the formation of charged domain walls in multiferroic $h$-$RE$MnO$_3$ is topologically inevitable because of the presence of highly curved vortex cores [8, 11, 12]. In this letter, we report the observation of distinct nanoscale conduction characteristics of charged ferroelectric domain walls in $h$-HoMnO$_3$ (a p-type semiconductor) using *in situ* conductive Atomic Force Microscopy (cAFM), piezo-response force microscopy (PFM), and kelvin-probe force microscopy (KPFM) at low temperatures. Local conduction spectra indicate that the conduction at tail-to-tail (TT) domain walls is significantly (slightly) enhanced at high forward (reverse) bias compared to that of the domains themselves, probably stemming from the accumulation hole-like carriers. In contrast, the conduction of head-to-head (HH) domain walls shows no enhancement at high forward bias and even suppression at high reverse bias, probably due to depletion of hole-like carriers.

---

[*] To whom correspondence should be address: wdwu@physics.rutgers.edu

$h$-$RE$MnO$_3$ are multiferroics with coexistence of ferroelectricity ($T_C \approx 1200 - 1500$ K) and antiferromagnetism ($T_N \approx 70 - 120$ K) [24, 25]. The ferroelectricity is induced by structural instability called trimerization [26, 27], where corner-shared MnO$_5$ polyhedra buckle to form Mn trimers with induced ferroelectric polarization along the $c$-axis. The uniaxial nature of the ferroelectricity allows only 180º domain walls. Combining three structural antiphase domain states ($\alpha$, $\beta$, $\gamma$) and two possible polarization states (+/-), there are totally 6 possible antiphase-ferroelectric domain states in $h$-$RE$MnO$_3$ ($\alpha^+$, $\beta^-$, $\gamma^+$, $\alpha^-$, $\beta^+$, $\gamma^-$). In 3-dimension, the six interlocked structural antiphase and ferroelectric domain walls merge to a line called vortex core, which is similar to vortex lines in type II superconductors. Experimentally curved vortex cores and domain walls along the $c$-axis were observed in $h$-$RE$MnO$_3$, forming charged ferroelectric domain walls [8, 11, 12].

Single crystals of $h$-HoMnO$_3$ were grown using the floating zone method. An atomically flat (110) surface was prepared by mechanical cleaving in ambient condition. The orientation of the $c$-axis is determined by Laué x-ray back scattering. The sample is glued to a sapphire substrate with Ag epoxy which serves as the back electrode. cAFM, PFM and KPFM experiments were carried out in a commercial variable temperature ultra-high vacuum (UHV) atomic force microscope (AFM) system [28]. PFM and cAFM were performed in contact mode with conductive AFM tip as the top electrode [10, 12]. A high performance current amplifier (noise level is less than 10 fA at 3 GΩ gain) was used for high sensitivity local current measurement. KPFM was carried out in tapping mode with ac bias modulated at 20-100 kHz. The AFM probes were either Pt/Ir or Au coated silicon cantilevers. The images were processed with a free software WSxM [29]. Samples for transmission electron microscopy (TEM) studies were prepared by Ar ion milling at liquid nitrogen temperature. A JEOL-2000FX and a JEOL-2010F TEM operating at 200 kV were used for the observations at room temperature.

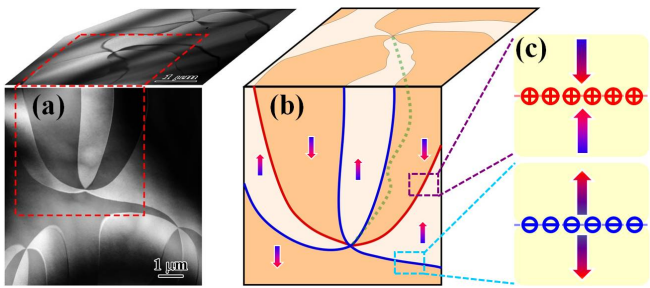

FIG. 1. (color online) (a) TEM dark field images of top and side view of vortex domains in *h*-HoMnO$_3$. (b) A cartoon sketch of 3D profile of a curved vortex in boxed area in (a). (c) Zoom-in cartoon of HH (TT) domain walls with positive (negative) bound charges.

Figure 1(a) shows typical room temperature TEM dark field images of *h*-HoMnO$_3$ taken with the electron beam either close to parallel (top view) or perpendicular (side view) to the crystallographic *c*-axis. The observation of vortex pattern on the side view image demonstrates that curved vortex cores force the associated domain walls to intercept the polarization direction (the *c*-axis), as illustrated in the cartoon in Fig. 1(b). Therefore, these domain walls carry bound charges due to antagonistic polarizations as illustrated in Fig.1 (c). In this sense, the charged domain walls in *h-RE*MnO$_3$ are "protected" by the formation of vortices, instead of being randomly pinned by extrinsic defects. In conventional ferroelectrics, charged domain walls are thermodynamically unstable because of extra electrostatic or strain energy cost [17]. However, *h-RE*MnO$_3$ are improper ferroelectrics where the ferroelectricity is a by-product of primary structural trimerization. Previous first principle calculations revealed that the ferroelectric order parameter (polarization) is a cubic function of the primary order near $T_C$ [27]. Therefore, the magnitude of the ferroelectric polarization is negligibly small in the proximity of $T_C$. In other words, the ferroelectric polarization and the associated electrostatic energy cost are irrelevant to the formation of the vortex-antivortex network near $T_C$. In addition, thermally-excited charge carriers can effectively screen charged domain walls because *h-RE*MnO$_3$ are small band gap semiconductors [9, 30]. From thermodynamic considerations, the proliferation of vortex-antivortex pairs and highly-curved vortex cores are more favorable because of the domination of the entropy contribution to the free energy at high temperature [1].

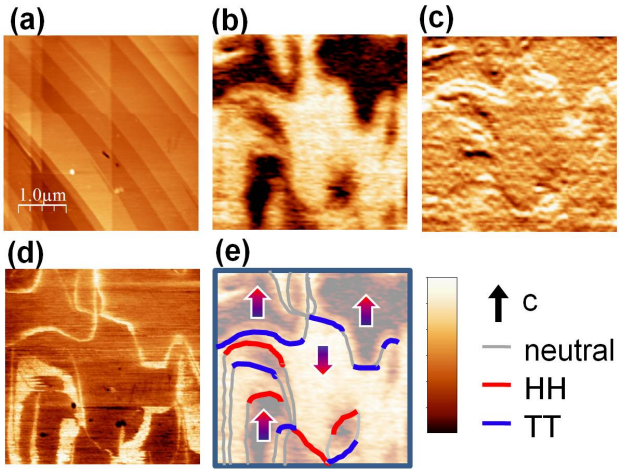

FIG. 2 (color online) (a) Topographic and (b), PFM ($V_{ex}$= 22 V, $f$=21 kHz) images taken simultaneously at 300 K. (c) derivative map of PFM image along the *c*-axis (vertical) where bright (dark) lines are TT (HH) domain walls. (d) cAFM image taken at the same location as PFM at 300 K with $V_{tip}$=-10 V. (e) A cartoon sketch of ferroelectric domain walls according to PFM (overlaid) and its derivative images. Red, blue and grey lines represent HH, TT and neutral domain walls, respectively. The arrows indicate in-plane polarization orientation determined from the phase of PFM signal. Color scales are 4.5 nm, 8 pm and 0.4 nA for topography, PFM and cAFM images respectively.

How vortices propagate along the *c*-axis can also be revealed by in-plane PFM with the *c*-axis in the surface plane [11, 12]. As seen in the topography image (shown in figure 2(a)), we are able to obtain an atomically flat (110) surface of a HoMnO$_3$ single crystal by mechanical cleaving. Fig. 2(b) shows the PFM images taken at room temperature to reveal ferroelectric domain pattern. We aligned the orientation of conductive cantilever so that it was parallel to the *c*-axis, which is along the slow scan axis (vertical direction). In this configuration, the PFM signal (vertical deflection) originates from the buckling of the cantilever caused by shear deformation of in-plane ferroelectric domains in the presence of an out-of-plane electric field [12, 31]. The dark and bright contrasts represent in-plane up and down ferroelectric domains, respectively. The direction of local polarization is determined by the phase of the PFM signal and is further confirmed by KPFM measurements at 65 K [32]. Fig. 2(c) shows the derivative image of PFM (b) image that highlights the HH (dark) and TT (bright) domain walls, respectively.

In cAFM images with $V_{tip}$=-10 V, there is no conduction (current) contrast between different domains, as shown in Fig. 2(d). In contrast, there are line features with significant extra current. More interestingly, these lines overlap with the ferroelectric domain walls observed by PFM at the same location. By correlating PFM and cAFM images at the same location, we can identify two vortices in this area, as illustrated in the cartoon sketch in Fig. 2(e). Note that the fast rastering of the cantilever artificially broadens the apparent width of some conduction lines. In the conductance image with very slow rastering (e.g. in Fig. 4), the observed width of conduction peaks is ~80-100 nm, which is comparable with the diameter ( ≤100 nm) of our conductive tip. To better illustrate the correlation between local conduction and ferroelectric domain walls, line profiles of PFM and cAFM images are plotted in Fig. 3(a). Clearly, significant enhancement (~25%) of current is observed at 180° TT domain walls. On the other hand, there is no current enhancement at HH domain walls. Interestingly, some conduction enhancement was observed in neutral domain walls that are running along the *c*-axis. The intriguing conduction enhancement at 180° TT domain walls has been confirmed in multiple locations of our single crystal sample using various AFM cantilevers with different conductive coating materials (see supplemental material), suggesting this is an intrinsic property of *h*-HoMnO$_3$.

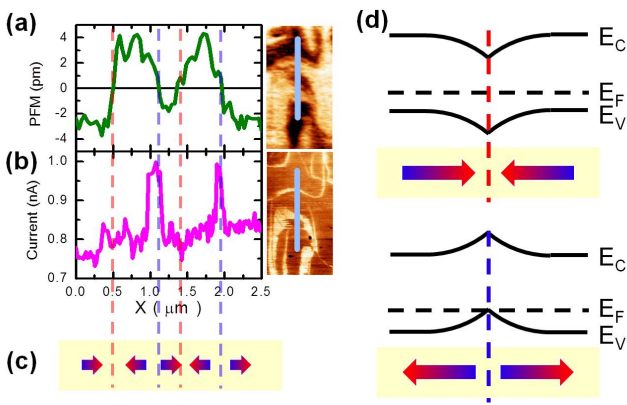

FIG. 3 (color online) (a) The PFM signal profile along the line on PFM image. (b) The current profile along the line on the current map at the same location. (c) A cartoon sketch of the corresponding ferroelectric domain configuration. (d) Schematic diagrams of band bending near charged domain walls. Upper panel: HH domain walls, lower panel: TT domain walls.

Previous cAFM studies on the (001) surface of $h$-HoMnO$_3$ suggest that the non-linear $I$-$V$ characteristics originate from a Schottky-like barrier between the tip (a metal) and crystal (a $p$-type semiconductor). In the (001) orientation, the polarization of the ferroelectric domains is either into or out of the surface plane; as a result, bound charges at the surface modulate the height of the Schottky-like barrier leading to an observable conduction contrast between opposite polarization domains in the cAFM images and $I$-$V$ curves [10]. On the (110) surface of HoMnO$_3$ explored in this work, the ferroelectric polarization is parallel to the surface and therefore there is no bound charge on the surface. Therefore, no conduction difference between opposite domains is expected at a given bias, which is consistent with lack of current contrast between opposite polarization domains in the cAFM images in Fig. 2(d) and Fig. 3(b). It has been predicted that charged 180º domain walls in ferroelectric semiconductors can attract free charge carriers with opposite sign, e.g. TT (HH) domain walls attract holes (electrons) [22, 23], and result in band bending. Since HoMnO$_3$ is a $p$-type semiconductor [33], the TT ferroelectric domain walls (which attract holes) should have enhanced conduction, while the HH domain walls (which repel holes) should have reduced conduction. This scenario, shown schematically in the band diagrams in Fig. 3(d), is qualitatively consistent with our observation.

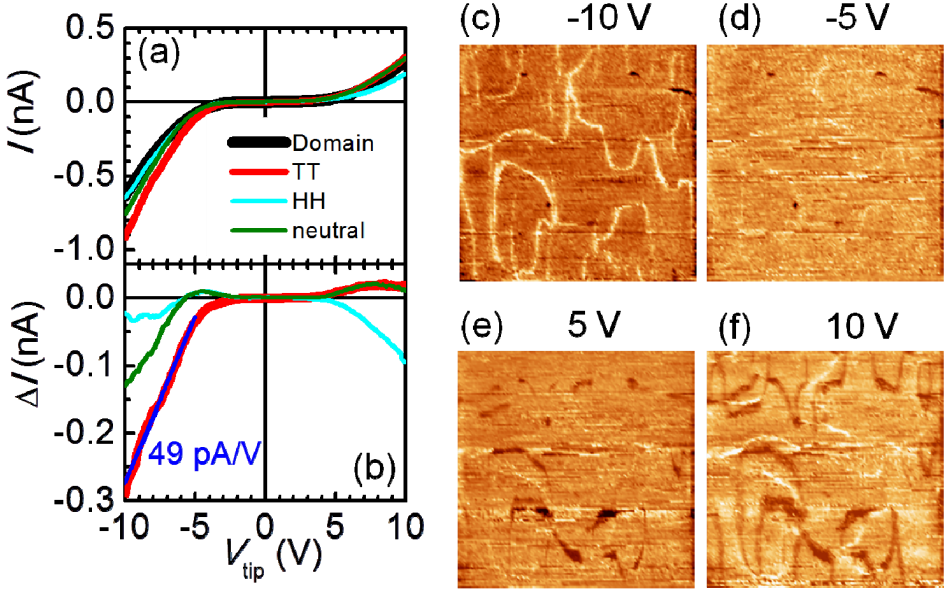

FIG. 4 (color online) (a) Areal averaged $I$-$V$ curves of ferroelectric domain (black), TT (red), HH (cyan) and neutral (green) domain walls. (b) $\Delta I$-$V$ curves of TT, HH and neutral domain walls after subtracting the areal average $I$-$V$ curve of domains. The blue line is a linear fit function (slope=49.0±0.7 pA/V) for -10 V<$V_{tip}$<-5V. (c)-(f) cAFM images of HoMnO$_3$ (110) surface taken with the spectro-microscopy mode. The biases (color scales) are -10 V (0.55 nA), -5 V (0.15 nA), 5 V (0.07 nA) and 10 V (0.35 nA), respectively. Brighter contrast corresponds to larger *absolute* current value.

In order to understand the conduction mechanism of charged domain walls in $h$-HoMnO$_3$, we performed $I$-$V$ spectro-microscopy measurements where an $I$-$V$ curve was measured at each pixel in the scanned area [10]. Areal average $I$-$V$ curves at domains, TT, HH and neutral domain walls are shown in Fig. 4(a), respectively. Consistent with previous results in ref. [10], the $I$-$V$ curves are nonlinear due to the Schottky-like barrier formed at tip-sample interface. The forward current at negative tip bias is consistent with the *p*-type character of the charge carriers in HoMnO$_3$ [10, 33]. Clearly, the TT domain wall is more conductive than the domains at high forward bias ($V_{tip}$<-5 V). The enhancement is smaller at reverse bias ($V_{tip}$>5 V). On the other hand, HH domain walls are less conductive than the domains themselves. To isolate the *I-V* character of the enhancement, we plotted the current difference ($\Delta I$) between the TT domain wall and the domain against the bias voltage ($V_{tip}$) (shown in Fig. 4(b)). Interestingly, the $\Delta I$-*V* curve shows an approximately linear increase (slope≈49 pA/V) beyond a forward bias $V_{tip}$<-5 V, which can be described by a linear fit function (the blue line in Fig. 4(b)): $\Delta I$=49×($V_{tip}$+4.45) (pA). We observed similar linear $\Delta I$-*V* characteristics (slope≈45 pA/V) at a different location with a different conductive AFM tip (see supplemental material), suggesting this enhancement is an intrinsic property of TT domain walls. The conduction enhancement indicates that there is a significant density of charge carriers in the proximity of TT domain walls, possibly forming two dimensional conducting sheets. On the other hand, there is no conduction enhancement at HH domain walls at high forward bias, while there is a clear suppression of conduction at high reverse bias, which is qualitatively consistent with a recent theory [23]. The neutral domain walls seem to be more conductive than domains at high forward bias ($V_{tip}$<-7 V), which may originate from a slight reduction of band gap due to the suppression of polarization at domain walls [13]. The systematic bias dependence of domain wall conduction enhancement (suppression) is shown in the current maps in Fig. 4(c) to (f) with bias voltages at -10 V, -5 V, 5 V and 10 V, respectively. Interestingly, both HH and TT domain walls are visible at reverse bias because of their antagonistic conduction behaviors.

The widths of HH domain walls (~120-250 nm) are larger than that of TT domain walls (~80-100 nm), suggesting larger screening length at positively charged domain walls. This is consistent with hole-type majority carriers because positively charged HH domain walls repel holes and generate a space charged region which increases the domain wall width [23]. Note that the observed width of TT domain walls may be limited by the cAFM tip radius (≤50 nm). Theoretically, the intrinsic width of charged 180º domain walls in proper ferroelectrics (e.g. PbTiO$_3$) is determined by the non-linear screening length because of significant accumulation of free charge carriers [22]. *h-RE*MnO$_3$ are improper ferroelectrics with very high $T_c$'s (>1000 ºC) and a sizable polarization (~5.5 μC/cm$^2$) at room temperature [25, 34]. In comparison, PbTiO$_3$ has lower $T_c$ (~490 ºC) but larger polarization (~75 μC/cm$^2$) [17]. In addition, the ferroelectric polarization of *h-RE*MnO$_3$ couples linearly to the primary order (trimerization) when *T* is sufficient lower than $T_c$ according to first principle studies [27], suggesting that the ferroelectricity of *h-RE*MnO$_3$ is effectively a proper one at room temperature. Therefore, it is reasonable to assume that the intrinsic width of TT domain walls in *h*-HoMnO$_3$ is comparable with the theoretical value in PbTO$_3$ (~2.7 nm) [23]. Future first principle and experimental studies may clarify this issue.

Transport of charge carriers confined at interfaces or surfaces is one of the major playgrounds in condensed matter physics for emergent phenomena, examples of which include the quantum hall effect at GaAs/AlGaAs with modulation doping [35], tunable superconductivity at LaAlO$_3$/SrTiO$_3$ with charge catastrophe [36] and multiferroic tunneling junction at the BaTiO$_3$/Fe interface [37]. However, many of these interfaces and surfaces suffer detrimental scattering of charge carriers from defects due to intrinsic imperfection in the fabrication process.

It has been speculated that charged ferroelectric domain walls, if naturally formed, may provide an alternate route for creating atomically clean interface for trapping charge carriers [23]. Yet most observed charged domain walls are metastable, often pinned by defects because they are energetically unfavorable [16-19, 21]. In multiferroic h-REMnO$_3$, charged ferroelectric domain walls form naturally because of the presence of topological defects, vortices. Utilizing *in situ* cAFM, PFM and KPFM we have observed nanoscale conduction enhancement at topologically-protected TT 180º domain walls in h-HoMnO$_3$. Our results suggest that this enhanced conduction originates from hole-like charge carriers that have accumulated to screen the negative bound charges. Our results demonstrate that topological defects can be harnessed to stabilize charged 180º domain walls in multiferroics, which opens up opportunities for a new kind of nanoscale conduction channel in multifunctional devices. Charged ferroelectric domain walls may provide novel platforms for creating correlated 2-dimensional electron gas without chemical doping.

We thank Mehmet Ramazanoglu for help crystal alignment with Laué x-ray scattering. W.W. thanks E. Lochocki and S. Park for optimizing PFM setup. We thank David Vanderbilt and Karin Rabe for helpful discussions and input. W.W. and S-W. C. acknowledge support from NSF DMR Award no. DMR-0844807 and no. DMR-1104484. Use of the Center for Nanoscale Materials was supported by the U. S. Department of Energy, Office of Science, Office of Basic Energy Sciences, under Contract No. DE-AC02-06CH11357.

# Supplemental Material for "Conduction of topologically-protected charged ferroelectric domain walls"

Weida Wu[1], Y. Horibe[1], N. Lee[1], S.-W. Cheong[1] and J.R. Guest[2]

[1]Department of Physics and Astronomy and Rutgers Center for Emergent Materials, Rutgers University, Piscataway, NJ 08854 USA
[2]Center for Nanoscale Materials, Argonne National Laboratory, Argonne, Illinois 60439 USA

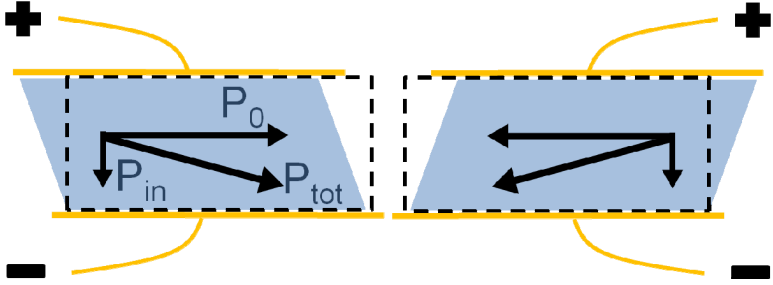

FIG. S1. Shear piezoelectric response of in-plane domains which determines the phase of the PFM signal at DC limit. The induced polarization causes rotation of total polarization, which in turn leads to the elongation of crystal along the diagonal direction, i.e. shear piezoelectric response.

**Kelvin-Probe Force Microscopy** (KPFM) is also known as Scanning Surface potential Microscopy, where a DC bias and an AC modulation with frequency ($\omega$) are applied to tip simultaneously (Eq. (1)). The AC component of resultant electric force between tip and sample surface provides a driving force at 1$^{st}$ and 2$^{nd}$ harmonic of the modulation (Eq. (2)), which cause the cantilever to oscillate at these frequencies. The amplitude of the 1$^{st}$ harmonic term linearly depends on the product of the AC modulation amplitude and the potential difference between tip and surface, as shown in Eq. (3). The DC component of electric force gradient causes the shift of resonant frequency which is a quadratic function of the potential difference, as shown in Eq. (4).

$$V_{tip} = V_{DC} + V_{AC} \cos(\omega t) \quad (1)$$

$$U = \frac{1}{2} C (V_{tip} - V_S)^2 \Rightarrow F_z = -\partial_z U = -\frac{1}{2} \partial_z C \cdot (V_{tip} - V_S)^2$$

$$F_z = -\frac{1}{2} \partial_z C \cdot \left[ (V_{DC} - V_S)^2 + 2(V_{DC} - V_S) \cdot V_{AC} \cos(\omega t) + V_{AC}^2 \cos^2(\omega t) \right] \quad (2)$$

The cantilever oscillation amplitude of 1$^{st}$ harmonic: $\quad Z_\omega \propto F_\omega \propto (V_{DC} - V_S) \cdot V_{AC} \quad (3)$

The resonant frequency shift ($\Delta f$) caused by force gradient: $\dfrac{\Delta f}{f_0} \propto \partial_z^2 F_z \propto (V_{DC} - V_S)^2$ (4)

We applied DC and AC bias ($V_{rms}$=5 V) to the conductive AFM tip in tapping mode. The 1st harmonic component was monitored with a lockin amplifier and the resonant frequency was tracked by a dedicated PLL. Eq. (3) and (4) are confirmed in our KPFM setup when ramping DC bias at a single spot, as shown in Fig. S2(d).

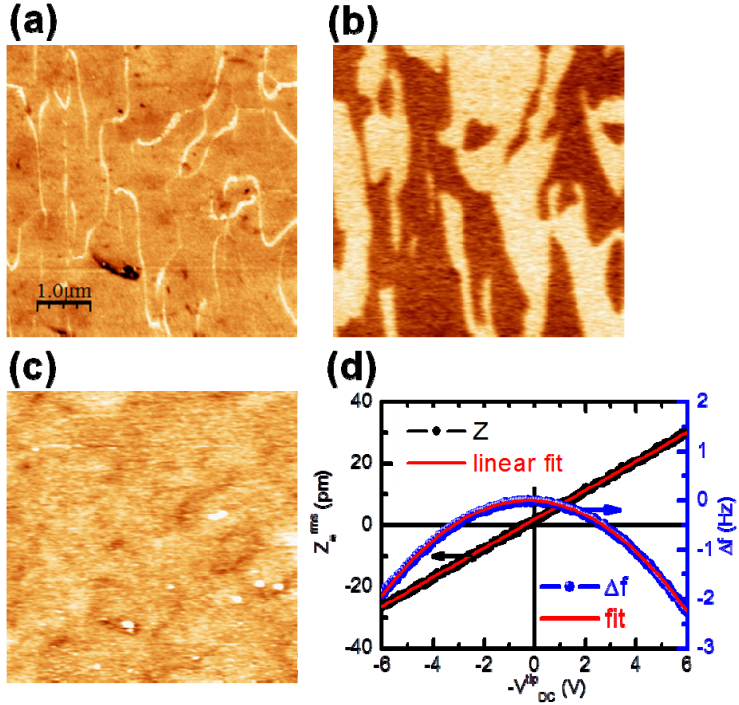

FIG S2. (a) cAFM image taken at RT with $V_{tip}$=-3V, (b) PFM image and (c) KPFM image taken at 65 K at the same location as the cAFM image. The dark contrast of KPFM image corresponds to more negative surface potential. Color scales are 400 nA, 10 pm and 1 V for cAFM, PFM and KPFM images, respectively. (d) Bias dependence of KPFM signal and frequency shift measured *simultaneously* at one spot of the sample surface with tapping mode. The red lines are fit functions according to theory in discussion.

We have carried out cAFM and PFM measurements with different conductive tips (Pt/Ir and Au) at different locations from the area show in main text. All the essential features were reproduced. One example is shown in Fig. S2. The cAFM image was taken at room temperature with $V_{tip}$=-3V. PFM and KPFM images were taken at 65 K. All images are taken at the same location. The darker contrast in KPFM image corresponds to more negative surface potential. The potential resolution is relatively poor in our setup because we were using stiff cantilevers with spring constant ~40 N/m for better PFM performance.

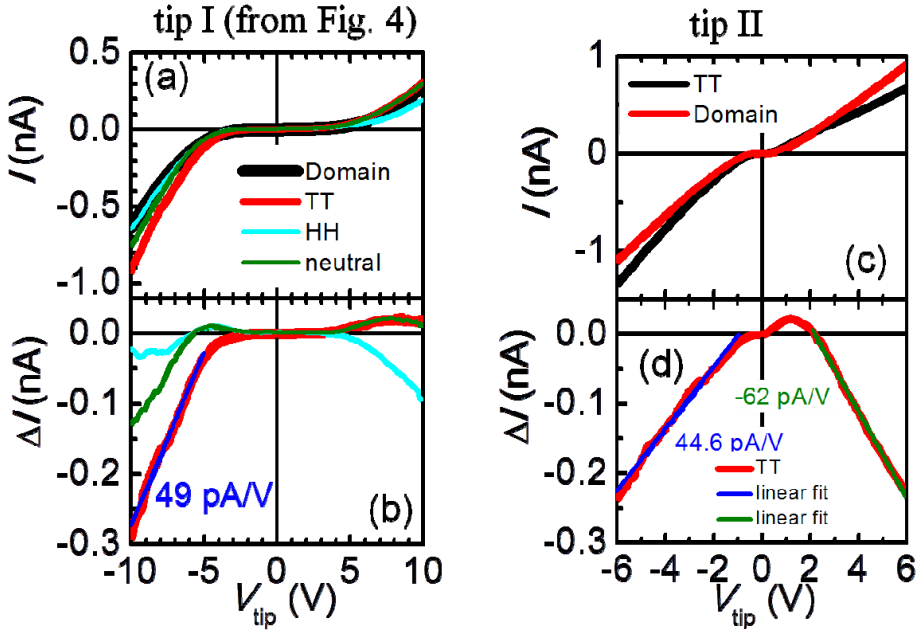

FIG. S3 (a) and (b) are areal average *I-V* curves and Δ*I-V* curves of head-to-head (HH), tail-to-tail (TT) and neutral domain walls, reproduced from Fig. 4 in main text. (c) and (d) are single point *I-V* curves and Δ*I-V* curves of head-to-head (HH) and tail-to-tail (TT) measured with a different cAFM tip at the location where Fig. S2(a) was taken.